\documentclass[Afour,sageh,times]{sagej}
\usepackage[utf8]{inputenc} 
\usepackage{moreverb,url}

\usepackage[colorlinks,bookmarksopen,bookmarksnumbered,citecolor=red,urlcolor=red]{hyperref}

\newcommand\BibTeX{{\rmfamily B\kern-.05em \textsc{i\kern-.025em b}\kern-.08em
T\kern-.1667em\lower.7ex\hbox{E}\kern-.125emX}}

\def\volumeyear{2016}

\begin{document}

\runninghead{Smith and Wittkopf}

\title{A Social Force Model for Companion Groups Considering Relative-Weight Attraction}

\author{Lihui Dong\affilnum{1} , Yingshuang He\affilnum{1} , Yunfeng Deng \affilnum{2} ,Qiuyu Zheng\affilnum{1} and Tianming Wang\affilnum{1}}

\affiliation{\affilnum{1}School of Safety Engineering, Shenyang Aerospace University, Shenyang, China\\
\affilnum{2}Party School of the Central Committee of C.P.C, National Academy of Governance, Beijing, China}

\corrauth{Yingshuang He,School of Safety Engineering, Shenyang Aerospace University, Shenyang, China.}

\email{308258543@qq.com}

\begin{abstract}
In response to the issue that traditional social force models do not adequately account for companion behavior with leaders and asymmetric information transmission during crowd evacuation simulations, this paper introduces an improved social force model for companion group that incorporates relative weight attraction. Based on the traditional social force model, the interaction forces among individuals within leader-follower groups are described by introducing relative weight attraction. Additionally, a velocity synchronization term is integrated into the pedestrians' self-driving force to address the problem of group dispersion commonly found in traditional models. Furthermore, the desired direction of followers within the companion group is refined to adapt to the evacuation movement led by the group leader. These enhancements collectively form a Social Force Model of companion groups considering relative weight attraction. Through comparative analysis of pedestrian evacuation processes in bidirectional channel simulation experiments between the relative weight attraction model and the traditional molecular potential (force) model, this study finds that the relative weight attraction model demonstrates stronger regulation and following capabilities when simulating collisions between companion groups and pedestrians or obstacles, more effectively ensuring group cohesion. Building upon this foundation, this study further investigates the impact of varying companion ratios on evacuation efficiency within the relative-weight attraction model. The results demonstrate that evacuation time steps exhibit a non-monotonic trend (initial increase, followed by a decrease, and a subsequent rise) as the companion ratio escalates. Additionally, across multiple simulation runs, the standard deviation of evacuation time steps expands with increasing companion ratios, indicating heightened fluctuation in individual evacuation timing. Model feasibility verification and case studies demonstrate that this model provides a theoretical foundation and a practical tool for simulating companion evacuation behavior with leaders and optimizing emergency evacuation strategies.
\end{abstract}


\maketitle

\section{1. Introduction}
With the acceleration of urbanization and the increasing frequency of large-scale public events, crowd gathering has become a common scenario in modern society that cannot be overlooked. The efficiency of emergency evacuation, the comfort of pedestrian flow under normal conditions, and the prevention of potential stampede risks in places such as shopping malls, stations, squares, and large event venues all rely on a profound understanding and accurate prediction of crowd movement patterns. As a theoretical framework that investigates the movement patterns of individuals within crowds, crowd dynamics provides a foundational basis for addressing the aforementioned issues. Among various microscopic simulation models, the Social Force Model \cite{R1}, renowned for its physical intuitiveness, behavioral interpretability, and effective characterization of complex interactions among individuals, has emerged as one of the predominant frameworks for describing and simulating crowd movement. The classical Helbing social force model and its subsequent developments have successfully replicated typical crowd phenomena such as "arching and clogging" and the "faster-is-slower" effect \cite{R2,R3,R4}, finding extensive applications in areas including evacuation planning and architectural design.

This assumption of the Social Force Model is reasonable for simulating physical collision avoidance, but it becomes overly simplistic when describing social interactions. Consequently, numerous scholars have dedicated efforts to improving the SFM, investigating its application in information transmission among individuals and between individuals and their environment, as well as exploring the incorporation of mechanisms for assigning leader roles. For instance, For instance, \cite{R5} investigated the impact of group effects on evacuation dynamics by implementing a novel model within the social force framework to explain social group behaviors in pedestrian crowds. \cite{R6} proposed a flexible model for representing social groups, extending the social force model to address both intra-group and inter-group interactions. Through simulation experiments, their study demonstrated how group structures (e.g., leader–follower relationships) influence the formation of "arching blockages." \cite{R7} developed a multi-exit evacuation model based on continuum mechanics as the dynamical foundation for pedestrian movement, guiding rational exit utilization in complex scenarios. \cite{R8} examined the effects of the quantity and positioning of evacuation leaders on crowd dynamics under limited visibility conditions. \cite{R9} incorporated pedestrian panic and its propagation mechanisms into an enhanced social force model to simulate crowd evacuation in subway transfer stations, providing valuable insights for metro operation management to understand collective evacuation behaviors. \cite{R10} explored how cultural differences influence crowd dynamics (e.g., evacuation efficiency, congestion formation) and quantified the role of cultural factors through an improved social force model. In the aforementioned models, most scholars have considered the utilization of the symmetric properties of attractive forces in social force models (short-range repulsion and long-range attraction). These symmetric models simplify pedestrian interactions into purely physical processes, thereby failing to adequately capture the asymmetric information transmission phenomena present in real-world companion groups. For instance, in groups with leaders, information flow between the leader and followers is asymmetric: the leader assumes a leading role based on evacuation information without necessarily attending to the followers, whereas the followers consistently monitor and respond to the leader’s movements during evacuation.

Based on the aforementioned issues, this paper establishes an asymmetric social force model with relative-weight attraction considering companion behavior (where A is attracted to B, but B is not attracted to A, hereafter referred to as the weighted model). In terms of companion behavior, an improved self-driving force and relative-weight attraction \cite{R11}  are introduced to characterize the attraction between companions, velocity coordination, and shared goal selection. Regarding guiding effects, by incorporating guide information and leadership-following mechanisms, the model investigates their influence on the decision-making of companion groups. Through the construction of an asymmetric social force model framework that couples "companion-leadership" dynamics, this study aims to reveal the movement patterns of companion groups in complex environments, thereby providing theoretical support for public space design, crowd management strategies, and risk prevention and control.

\section{2. Basic Model}
In 1995, \cite{R1} proposed the Social Force Model to describe pedestrian motion, based on Newton’s second law. This model incorporates several key forces: the self-driving force guiding pedestrians toward their desired velocity, interaction forces between pedestrians, and interaction forces with obstacles such as walls. The system is formulated through a set of partial differential equations as follows:
\begin{equation}
m_{i} \frac{d v_{i}(t)}{d t}=f_{i}^{0}+\sum_{j( \neq i)} f_{i j}+\sum_{w} f_{i w}
\end{equation}

\begin{equation}
f_{i}^{0}=m_{i} \frac{v_{i}^{0} e_{i}^{0}\left(t\right)-v_{i}(t)}{\tau}
\end{equation}

\begin{equation}
\begin{aligned}
& f_{ij}=\left\{A\exp\left[\frac{D - d_{ij}}{B}\right]\right\}n_{ij}+K\Theta\left(D - d_{ij}\right)n_{ij}\\
& +K\Theta\left(D - d_{ij}\right)\Delta v_{ji}t_{ij} \\
\end{aligned}
\end{equation}

\begin{equation}
\begin{aligned}
& f_{iw}=\left\{A \exp \left[\frac{r - d_{iw}}{B}\right]\right\} n_{iw}+K \Theta\left(r - d_{iw}\right) n_{iw}\\
& +K \Theta\left(D - d_{iw}\right)\left(v_{i} \cdot t_{iw}\right) t_{iw} \\
\end{aligned}
\end{equation}

The equation indicates that the social force acting on pedestrian i consists of three distinct components: the self-driving force, which embodies the pedestrian's conscious movement intention and can be interpreted as the attraction toward their target destination; the psychological repulsion force between pedestrian i and other pedestrians, reflecting the extent of avoidance during movement to prevent physical contact or collision; and the interaction force between pedestrian i and obstacles such as walls.

Research indicates that the Social Force Model can effectively reproduce self-organization phenomena in crowds. Subsequently, scholars have explored more complex scenarios by modifying the forms of interpersonal forces to better align with realistic pedestrian evacuation dynamics. For instance, \cite{R12} represented pedestrians using three intersecting circles rather than a single circle, introducing rotational effects into the model to simulate high-density crowd movement. \cite{R13} extended the Social Force Model to incorporate overtaking behavior during pedestrian motion, wherein individuals with higher desired speeds catch up and pass those moving in the same direction at lower desired speeds. Furthermore, \cite{R14} employed a three-circle representation to simulate rotational movements and expanded the Social Force Model to characterize pedestrian dynamics on staircases. \cite{R15} quantified pedestrians' physiological and psychological attributes by introducing a physical constitution coefficient and a mental state coefficient, thereby establishing a social force model based on pedestrian heterogeneity. \cite{R16} incorporated an information transmission mechanism into the social force model to simulate pedestrian behavior under emergency conditions.

\section{3. Establishment of a Social Force Model for Companion Groups Considering Relative-Weight Attraction }
\subsection{3.1. Improvement of Pedestrian Self-Driving Force}
During the evacuation process of companion groups, the self-driving force within the Social Force Model encompasses both the self-driving force of individuals evacuating alone and that of the companion group as a collective whole. The individual’s self-driving force reflects their instinctive response to avoid danger and seek safety, whereas the collective self-driving force of the companion group stems from mutual support and coordination among members, including velocity adaptation. This dual self-driving mechanism enables individuals to maintain their own safety during emergencies while simultaneously relying on companion relationships for psychological support and guidance. Such interaction not only influences individual speed and movement decisions but also enhances the overall efficiency and safety of the evacuation process.

\begin{figure}[h]
\centering
\includegraphics[scale=0.6]{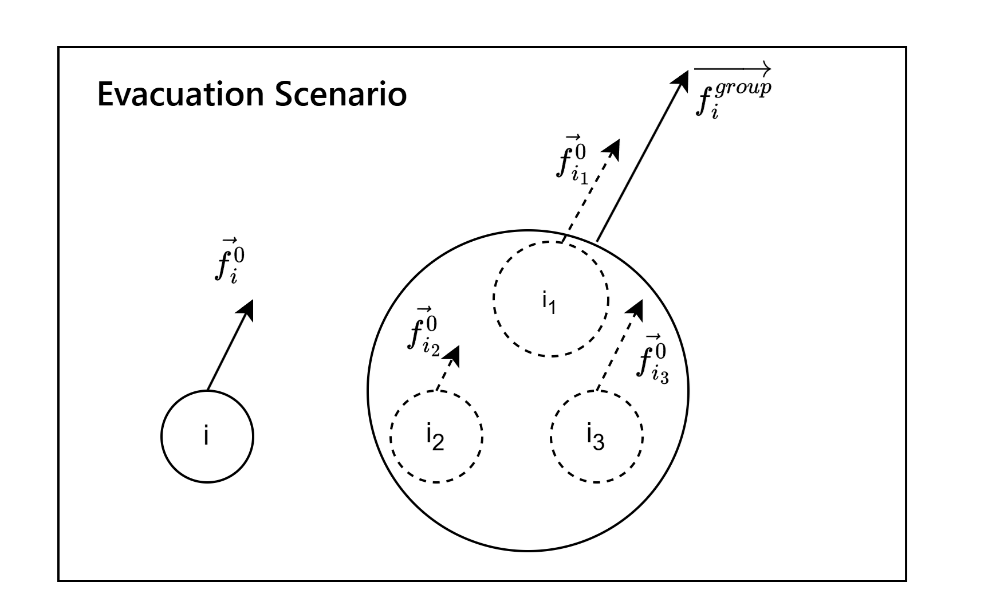}        
\caption{Schematic diagram illustrating the self-driving forces present in companion groups.}
\label{fig2} 
\end{figure}

Therefore, the enhancement of the self-driving force for companion groups can be approached from the following aspects:

At the psychological level, group members exhibit a tendency to maintain similar velocities to avoid separation from the collective. A velocity synchronization term is incorporated into the self-driving force to model this behavior.

\begin{equation}
{\bf{f}}_i^{{\rm{desire }}} = \underbrace {{m_i}\frac{{{\bf{v}}_i^{0{\rm{ }}}{e_i}^0\left( t \right) - {{\bf{v}}_i}}}{\tau }}_{{\bf{Basic}}{\rm{ }}{\bf{driving}}{\rm{ }}{\bf{force}}} + \underbrace {\varepsilon \sum\limits_{k \in G} {\left( {{v_k} - {v_i}} \right)} }_{{\bf{velocity}}{\rm{ }}{\bf{synchronization}}{\rm{ }}{\bf{term}}}
\end{equation}

In this formula, $\varepsilon$ represents the velocity synchronization coefficient, where a larger value indicates stronger velocity coordination within the group; $\varepsilon\sum\limits_{k\in G} {\left({{{\bf{v}}_k} - {{\bf{v}}_i}}\right)}$ denotes the velocity synchronization mechanism; and $\sum\limits_{k \in G} {\left( {{{\bf{v}}_k} - {{\bf{v}}_i}} \right)} $ signifies the difference between the individual’s velocity and the group’s average velocity.

The coefficient $\varepsilon$ can be determined by referring to the velocity synchronization coefficient ${p_{group}}(t)$   introduced in the study by \cite{R17}. In their model, this parameter is defined as the average modulus of the direction vectors of all individuals, with a value range of [0, 1]. A value closer to 1 indicates high synchronization strength, reflecting highly consistent directional movement among individuals, whereas a value approaching 0 signifies low synchronization strength, representing random directional behavior.

Furthermore, during personnel evacuation, interactions within companion groups are profoundly influenced by their emotional states. Especially in emergencies, different groups exhibit varying levels of panic due to physiological and psychological differences. These emotional disparities not only affect their judgment and behavioral choices but also significantly impact the overall efficiency of the evacuation process. Under these circumstances, the degree of dependency among individuals during evacuation can be quantified using pedestrians' emotional values, providing a critical theoretical basis for understanding and optimizing evacuation procedures. Therefore, when modeling the self-driving force of companion pedestrians, panic psychology must be incorporated. During evacuation, emotional values influence pedestrians' desired speed \cite{R18}. Accordingly, the actual speed of pedestrians under different emotional values can be expressed by Equation (8).

\begin{equation}
{v_i}(t) = {v_{\min }}(1 - {E_i}(t)) + {E_i}(t) \cdot {v_{\max }}
\end{equation}

In this formula, ${v_i}(t)$represents the actual walking velocity attained by individual i at time t, which is influenced by the current emotional value—the greater the panic, the closer the actual velocity approaches the maximum velocity. ${v_{\min }}$ denotes the minimum velocity of the individual, while ${v_{\max }}$ signifies the maximum achievable velocity. ${E_i}(t)$ corresponds to the emotional value of pedestrian i at time t. Notably, ${v_{\max }}$ is not a fixed parameter; it varies based on the pedestrian’s gender and age\cite{R19}. The specific emotional values are detailed in Table 1 \cite{R20}.

\begin{table}[h]
\small\sf\centering
\caption{Classification of Emotional States.\label{T1}}
\begin{tabular}{lll}
\toprule
Emotion Value&State&Behavioral Characteristics\\
\midrule
\texttt{0}&Immune (Black)&Unaffected by Emotion\\
\texttt{(0,0.3]}&Calm (Green)&Rational Decision-Making\\
\texttt{(0.3,0.7]}&Anxious (Yellow)&Impaired Judgment\\
\texttt{((0.7,1]]}&Panic (Red)&Loss of Behavioral Control\\
\bottomrule
\end{tabular}
\end{table}

The calculation of self-driving forces for different pedestrian groups must account for their individual mass and initial velocity. To address this, researchers have classified pedestrian velocities into ten distinct categories \cite{R21} , thereby enabling the computation of the modified self-driving force for companion pedestrians.

\subsection{3.2. Relative-Weight Attraction Model for Companion Pedestrians}
The interaction forces among companion pedestrians, when considering the group as a cohesive unit, involve multiple physical parameters, including individual body size, movement radius, friction, and compression forces. Collectively, these factors shape the interpersonal relationships within companion groups and influence their behavioral patterns and mobility during evacuation. Pedestrian interactions can be categorized into repulsive forces (for collision avoidance) and attractive forces (stemming from social bonds). During companion walking, repulsive forces between individuals are often negligible. In traditional models, researchers have employed intermolecular potential (forces) to simulate short-range repulsion and long-range attraction among pedestrians[5] , under the assumption that attraction is symmetric and globally normalized. However, in groups with leaders, this approach fails to address inherent defects in modeling asymmetric attraction.

To address this, a relative-weight attraction model for companion groups can be established by introducing the relative-weight attraction . This approach redefines the attraction between companions as a relative-weight attraction generated by their mutual influence weights. The model is particularly suitable for asymmetric crowd structures with leaders, as shown in Equation (9).

\begin{equation}
F_i^{{\rm{attr}}} = \alpha  \cdot \frac{{\sum\limits_j \phi  (|{x_j} - {x_i}|) \cdot ({x_j} - {x_i})}}{{\sum\limits_k \phi  (|{x_k} - {x_i}|)}}
\end{equation}

In this formulation, $\alpha $ denotes the attraction intensity, which governs the magnitude of attractive acceleration; $\phi (r)\;$  represents the companion influence function; and $r = \parallel {x_j} - {x_i}\parallel $ signifies the Euclidean distance between companions. The incorporation of relative-weight attraction into the Social Force Model offers a locally adaptive advantage: the attractive force depends solely on the local distribution density of surrounding companions rather than on the global population size N. This is achieved through the use of an active set that filters and identifies effective companions, defined formally in Equation (10).

\begin{equation}
{\Lambda _i}(\theta ) = \left\{ {j\left| {\phi \left( {\left| {{x_j} - {x_i}} \right|} \right)} \right. \ge \theta  \cdot {{\max }_k}\phi \left( {\left| {{x_k} - {x_i}} \right|} \right)} \right\}
\end{equation}

Compute the attraction force exclusively among companions within set $\Lambda i(\theta )$.

\begin{equation}
F_i^{{\rm{attr}}} = \alpha  \cdot \frac{{\sum\limits_{j \in {\Lambda _i}} \phi  ({r_{ij}})({x_j} - {x_i})}}{{\sum\limits_{k \in {\Lambda _i}} \phi  ({r_{ik}}) + \rho }}
\end{equation}

In this formula, $\rho $ represents a minuscule constant to prevent division by zero, $\rho  = {10^{ - 5}}$; $\phi (r)\;$ denotes the companion influence function, which can be represented by function $\phi (r) = {(1 + {r^2})^{ - 1}}$ characterized by computational simplicity and smoothness, adhering to the principle of distance decay.

Based on the aforementioned improvements to the traditional social force model, evacuation simulations for asymmetric companion pedestrians with leaders can be effectively conducted. In such groups, no equilibrium distance exists between members, meaning individuals do not spontaneously maintain fixed spacing. The attraction between pedestrians depends solely on the distribution of effective surrounding companions (independent of distant crowds). This asymmetric interaction force breaks the action-reaction symmetry inherent in traditional mechanics, better aligning with real-world social scenarios involving leaders. Consequently, the model more accurately captures leader-follower behaviors, information transmission pathways, and group decision-making processes.

\subsection{3.3. Desired Direction of Motion for Companion Pedestrians}
In companion groups with leaders, the bellwether-follower mechanism with age-based role allocation can be implemented. Scholarly research indicates that leader selection primarily depends on proximity to exits; however, when age grouping is introduced, leadership roles may be determined by age hierarchy (e.g., older individuals may proactively guide children even if marginally farther from the exit). Therefore, it is necessary to adjust the leader determination logic by integrating exit proximity with age grouping and priority rules. The formal criterion for leader identification is given in Equation (12).

\begin{equation}
{\rm{Leader}} = \left\{ {\begin{array}{*{20}{l}}
{\arg {{\min }_{z \in {G_A}}}{d_{iE}}}&{{\rm{if }}{G_A} \ne 0}\\
{\arg {{\min }_{z \in {G_E}}}{d_{iE}}}&{{\rm{else if }}{G_E} \ne 0}\\
{\arg {{\min }_{z \in {G_C}}}{d_{iE}}}&{{\rm{otherwise}}}
\end{array}} \right.
\end{equation}

In this formula, ${G_A}$ represents members of the adult subgroup; ${G_E}$ denotes members of the elderly subgroup; ${G_C}$ signifies members of the child subgroup; ${d_{iE}}$ indicates the distance to the exit, $\left\|\vec{P}_{i}-\vec{P}_{\text{exit}}\right\|$; ${\vec P_i}$ corresponds to the member's positional coordinates, $(x_i, y_i)$ ; and ${\vec P_{exit}}$ represents the exit's positional coordinates, $(x_{exit}, y_{exit})$. The rule for determining the leader is defined as follows: the priority rule across age groups is adults $>$  elderly $>$ children. Within the same age group, the individual closest to the exit is selected as the leader. The desired direction of the leader is given by Equation (13).

\begin{equation}
{\vec e_L} = (1 - \mu ){\vec e_{LE}} + \mu {\vec e_a}
\end{equation}

In this formula, ${\vec e_{LE}}$ represents the unit vector pointing from the leader’s current position to the exit; ${\vec e_a}$ denotes the unit vector directing toward the group center (cohesion direction); $\mu$ signifies the exit direction weighting coefficient, reflecting the extent to which the leader prioritizes the exit direction over group influence.

The exit direction weighting coefficient $\mu$ quantifies the influence of group dynamics on the leader’s desired direction. A smaller value indicates that the leader’s directional decision is less affected by the companion group, emphasizing instead self-driven evacuation behavior. The desired direction for followers is explicitly defined in Equation (14).

\begin{equation}
{\vec e_i} = {\beta _i}{\vec e_L} + {\gamma _i}{\vec e_{iL}} + {\delta _i}{\vec e_{iE}}
\end{equation}

In this formulation, ${\beta _i}$ denotes the follow-the-leader weight; ${\gamma _i}$ represents the group cohesion weight; $[{\delta _i}$ signifies the autonomous evacuation weight; ${\vec e_L}$ corresponds to the leader’s desired direction; ${\vec e_{i{\rm{L}}}}$ indicates the unit vector from follower i to the leader; ${\vec e_{{\rm{i}}E}}$  refers to the unit vector from follower i to the exit. The schematic diagram of the desired movement direction for companion pedestrians is illustrated in Figure 2.

\begin{figure*}[h]
\centering
\includegraphics[scale=0.5]{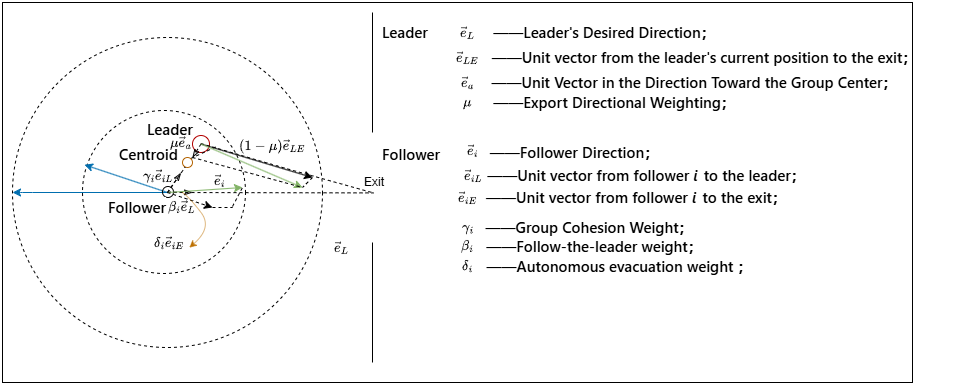}        
\caption{Schematic Diagram of the Desired Direction of Movement for Companion Pedestrians with Leaders.}
\label{fig2} 
\end{figure*}

Both leaders and followers exhibit desired directions of movement toward exits and toward group cohesion. By incorporating an age-stratified role-based decision-making mechanism, the model captures adults' autonomous decision-making power (reflecting realistic cognitive capabilities), children's high followership dependency (aligning with characteristic reliance behaviors), and group-centric attraction (embodying the core concept of group cohesion). Through optimizing the desired direction of companion groups, this framework more accurately simulates real-world collective evacuation behavior. It simultaneously accounts for intra-group interactions within compact clusters, integrates an optimized bellwether mechanism, and incorporates behavioral characteristics across age groups, thereby providing a theoretical foundation for expectation orientations in multi-age companion dynamics.

\section{4. Simulation Analysis of Companion Groups Utilizing Relative Weight Attraction}
\subsection{4.1. Selection of Experimental Parameters}
Individuals tend to form small groups of 2–3 people during evacuation processes. Therefore, a group size of 2–3 individuals was selected for this study to model companion behavior. The simulation environment is configured as a two-way passage spanning –25 m to 25 m along the x-axis and –20 m to 20 m along the y-axis, creating a typical pedestrian simulation context. Obstacles are strategically placed to influence pedestrians’ path selection, forcing detours or reduced speeds, thereby increasing evacuation complexity. At the initial timestep, pedestrians are distributed according to a predefined pattern. The simulation is implemented using Python, building upon the PySocialForce open-source project. Key enhancements include the incorporation of a leader assignment mechanism, desired direction improvements, and relative-weight attraction functions to simulate companion group dynamics under the proposed model.

\subsection{4.2. Simulation Scheme}
Three simulation schemes are designed: Scheme 1: Generate companion groups and simulate collisions with other companion groups and obstacles, comparing the traditional social force model utilizing molecular potential (force) to validate its feasibility in simulating companion group dynamics; Scheme 2: Contrast the traditional social force model employing molecular potential (force) to examine the impact of leader presence on evacuation efficiency, analyzing the advantages and limitations of the relative-weight model; Scheme 3: Compare the effects of six companion ratios (0\%, 30\%, 50\%, 70\%, 90\%, and 100\%) on evacuation efficiency under the relative-weight attraction model. The experimental schemes are summarized in Tables 2-4. To ensure consistency in controlled variables and code simplicity across model comparisons, the velocity distribution mechanism from the PySocialForce open-source project is adopted for speed parameterization.

\begin{table*}[h]
\small\sf\centering
\caption{Experimental Scheme 1.\label{T1}}
\resizebox{\textwidth}{!}{
\begin{tabular}{lllllll}
\toprule
\multicolumn{2}{c}{Group}&Number of people&Direction of motion&Presence or Absence of a Leader&Presence or Absence of Obstacles&Number of Simulation Runs\\
\midrule
\texttt{Traditional Social Force Model}&Small group&2&From left to right&Absence&Absence&25\\
\texttt{}&Large group&5&From right to left&Absence&Absence&25\\
\texttt{}&Obstacle collision&3&From left to right&Presence&Absence&25\\
\texttt{Relative-Weight Attraction Mode}&Small group&2&From left to right&Presence&Absence&25\\
\texttt{}&Large group&5&From right to left&Presence&Absence&25\\
\texttt{}&Obstacle collision&3&From left to right&Presence&Presence&25\\
\bottomrule
\end{tabular}}
\end{table*}

Scheme 1: Design a collision scenario between small and large groups to validate the rationality of pedestrian-pedestrian interactions. In a two-way passage scenario, two groups are generated: a small group (2 agents) and a large group (5 agents). The small group is initially positioned on the left side of the passage and moves rightward, while the large group starts from the right side and moves leftward. Both groups converge and collide in the central section of the passage, creating interactive interference. The velocities (in both x and y directions) of the small group members (ID=0 and ID=1) are recorded. To mitigate data randomness, multiple simulation iterations are conducted.

\begin{figure*}[h]
\centering
\includegraphics[scale=1]{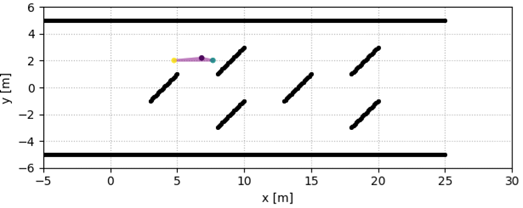}        
\caption{Collision between Pedestrians and Obstacles.}
\label{fig2} 
\end{figure*}

Design a collision scenario between a small group and obstacles to validate the rationality of pedestrian-environment interactions. The group (3 agents) evacuates from left to right toward the exit. Multiple obstacles approximately 1 meter in size are positioned along the evacuation path, as illustrated in Figure 3, to influence pedestrian route selection. The velocity components along the x and y axes for all group members (ID=0, ID=1, and ID=2) are recorded. To mitigate data randomness, multiple simulation iterations are conducted.

\begin{table*}[h]
\small\sf\centering
\caption{Experimental Scheme 2.\label{T1}}
\resizebox{\textwidth}{!}{
\begin{tabular}{lllll}
\toprule
Group&Companion Ratio(Left-Side)&Presence or Absence of a Leader&Presence or Absence of Obstacles&Number of Simulation Runs\\
\midrule
\texttt{Traditional Social Force Model}&0&Presence&Presence&25\\
\texttt{}&50\%&Presence&Presence&25\\
\texttt{}&90\%&Presence&Presence&25\\
\texttt{Relative-Weight Attraction Model}&0\%&Absence&Presence&25\\
\texttt{}&50\%&Absence&Presence&25\\
\texttt{}&90\%&Absence&Presence&25\\
\bottomrule
\end{tabular}}
\end{table*}

Scheme 2: Simulations of pedestrian movement with three distinct companion ratios are conducted in a bidirectional passage for total populations of 60, 90, 120, 150, 180, 210, and 240 individuals. The relative-weight model (with leaders) and the traditional molecular potential (force) model (without leaders) are employed to simulate evacuation processes, with evacuation efficiency statistically analyzed for the right exit. Multiple obstacles are incorporated to increase evacuation complexity, testing the impact of leadership roles on evacuation efficiency under constrained conditions. To mitigate data randomness, results are averaged over 25 independent simulation runs. Furthermore, to avoid over-reliance on the 100\% evacuation completion time (which may be skewed by outliers such as extremely slow individuals or accidental blockages), the time required for 90\% of the population to evacuate is recorded, reflecting the flow efficiency of the majority.

\begin{table*}[h]
\small\sf\centering
\caption{Experimental Scheme 3.\label{T1}}
\resizebox{\textwidth}{!}{
\begin{tabular}{lllll}
\toprule
Companion Ratio&Number of people&Presence or Absence of a Leader&Presence or Absence of Obstacles&Number of Simulation Runs\\
\midrule
\texttt{0}&1500&Presence&Presence&25\\
\texttt{30\%}&150&Presence&Presence&25\\
\texttt{50\%}&150&Presence&Presence&25\\
\texttt{70\%}&150&Presence&Presence&25\\
\texttt{90\%}&150&Presence&Presence&25\\
\texttt{100\%}&150&Presence&Presence&25\\
\bottomrule
\end{tabular}}
\end{table*}

Scheme 3: In a bidirectional passage, simulations of pedestrian movement with six distinct companion ratios are conducted for a total population of 150 individuals using the relative-weight model. The results are averaged over 25 independent simulation runs, with the evacuation completion time defined as the point at which 90\% of the population has evacuated. Obstacles are appropriately incorporated to enhance scenario realism.

\subsection{4.3. Feasibility Comparison and Validation of the Relative-Weight Attraction Model}
In pedestrian evacuation simulations, a comparative analysis between the relative-weight attraction model and the traditional social force model utilizing molecular potential (force) was conducted to validate the feasibility of the weighted model in evacuating pedestrians. Using Experimental Scheme 1 for simulation, the velocity simulation line results for small group members (ID=0, 1) and the post-interaction velocity simulation line results between pedestrians and the environment are illustrated in Figure 4-5.

\begin{figure*}[h]
\centering
\includegraphics[scale=1]{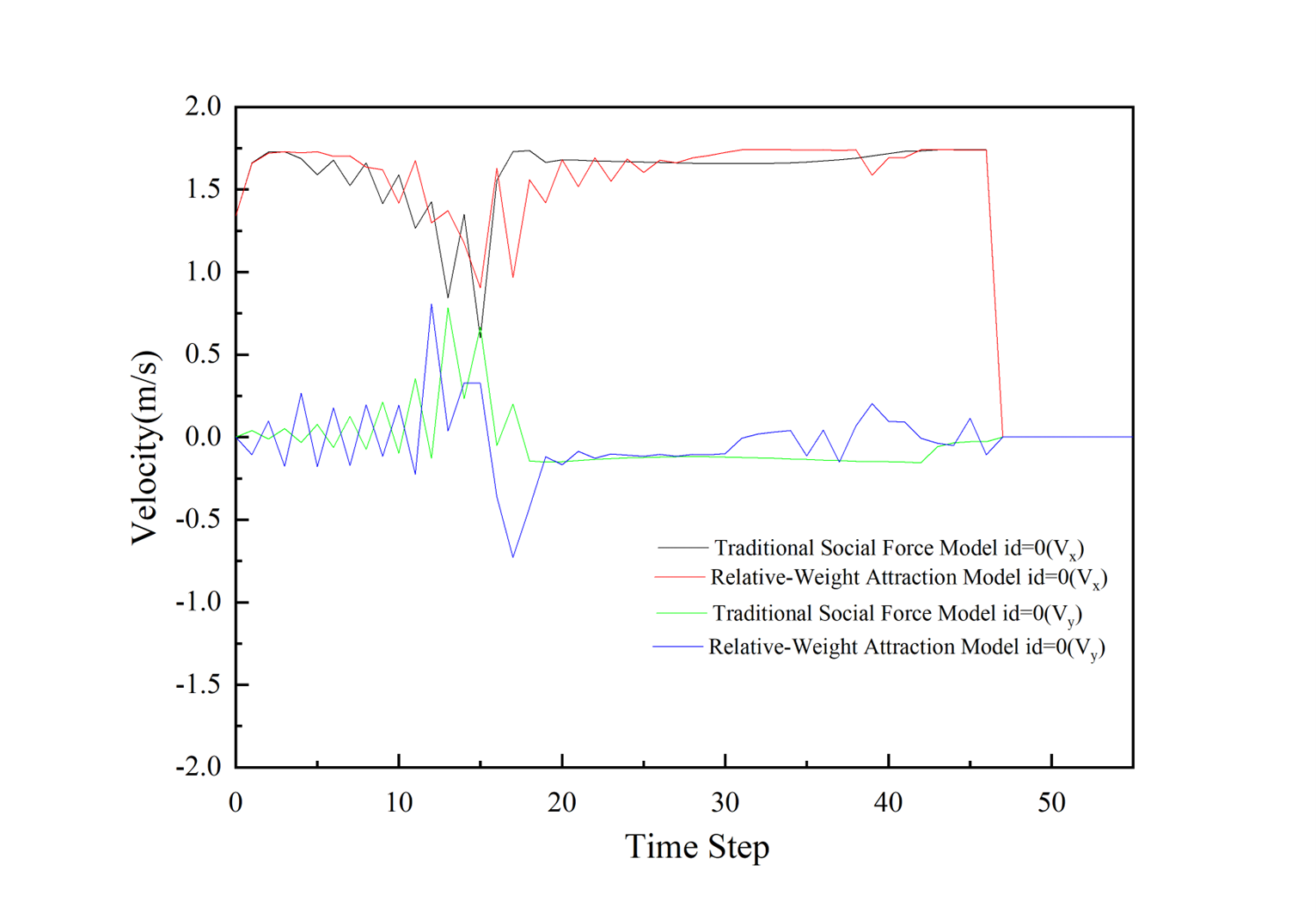}        
\caption{Velocity Simulation Line Result Chart for Small Group Member (ID=0) .}
\label{fig2} 
\end{figure*}

\begin{figure*}[h]
\centering
\includegraphics[scale=1]{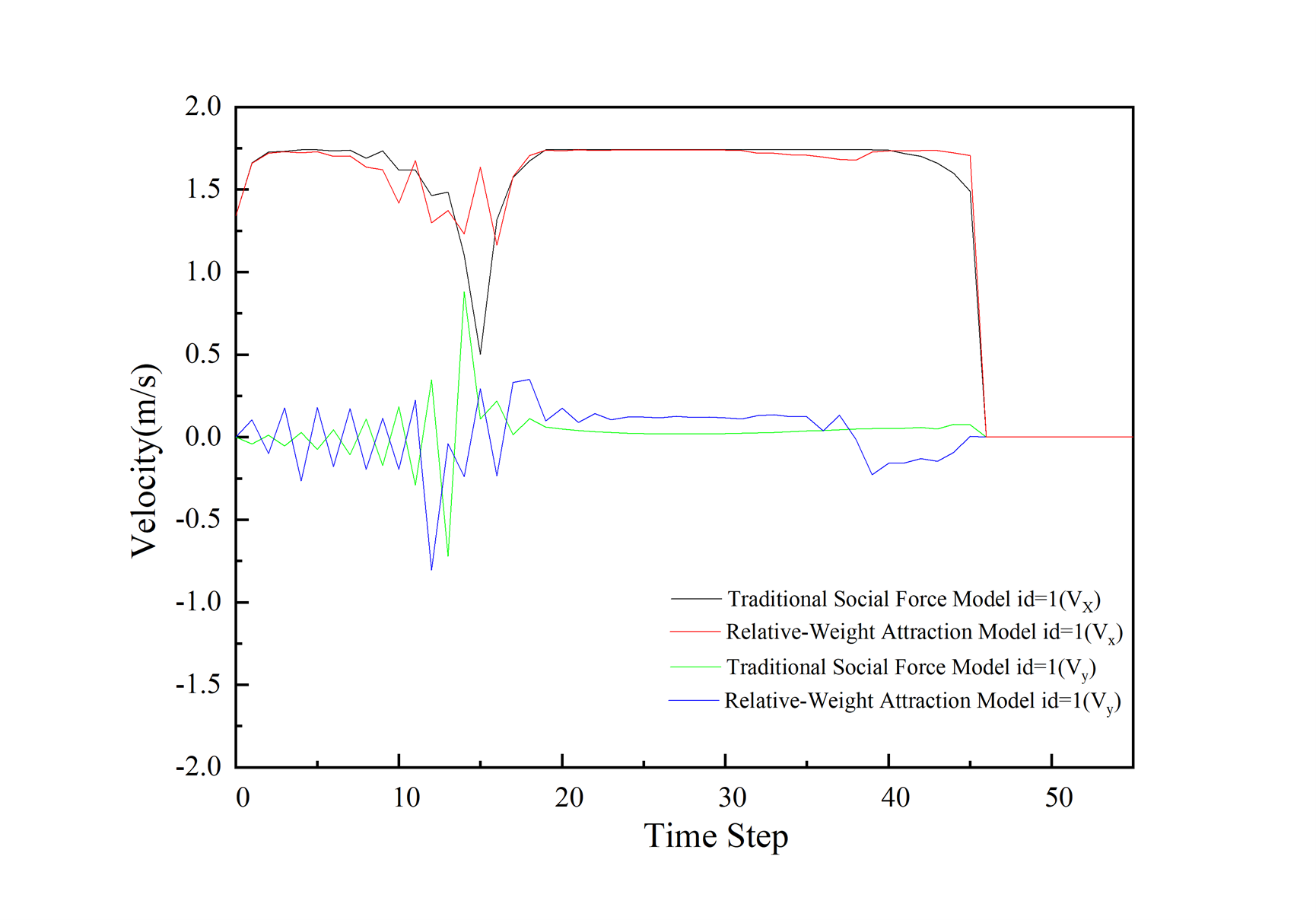}        
\caption{Velocity Simulation Line Result Chart for Small Group Member (ID=1) .}
\label{fig2} 
\end{figure*}

From the velocity profile line charts in crowd collision simulations, it is evident that both models can capture the fundamental behavioral characteristics of companion group evacuation: 1. Consistent Velocity Trends: Both the traditional and weighted models exhibit similar velocity trends for agents ID=0 and ID=1 within the time step range of 0–50. After initial fluctuations, velocities stabilize and eventually approach zero around time step 50, indicating that both models successfully simulate pedestrians reaching their target locations. 2. Group Coordination Dynamics: Both models demonstrate coordination among group members, particularly during the velocity fluctuation phase between time steps 10–20. This reflects the models’ capacity to simulate mutual influence and synchronized behavior within companion groups during evacuation
In the weighted model, the Vx velocity of ID=1 (designated as the leader) remains relatively stable throughout the process (approximately 1.5 m/s), whereas in the traditional model, the Vx velocity of ID=1 exhibits significant fluctuations. This demonstrates the leader’s stable guiding role in the weighted model. Meanwhile, the Vy velocity of ID=0 (designated as the follower) in the weighted model shows more pronounced fluctuations, particularly during time steps 10–20 and 30–50, indicating that the follower adjusts its movement speed based on the leader’s position. These fluctuations reflect the capability of the relative-weight attraction model to simulate the follower’s dynamic behavioral adjustments in response to the leader, thereby better aligning with the asymmetric interaction patterns observed in real-world companion groups.
Simulations are conducted using the three-agent companion group defined in Experimental Scheme 1. The velocity simulation line chart results for pedestrians following obstacle collisions are illustrated in Figure 6-9.

\begin{figure*}[h]
\centering
\includegraphics[scale=1]{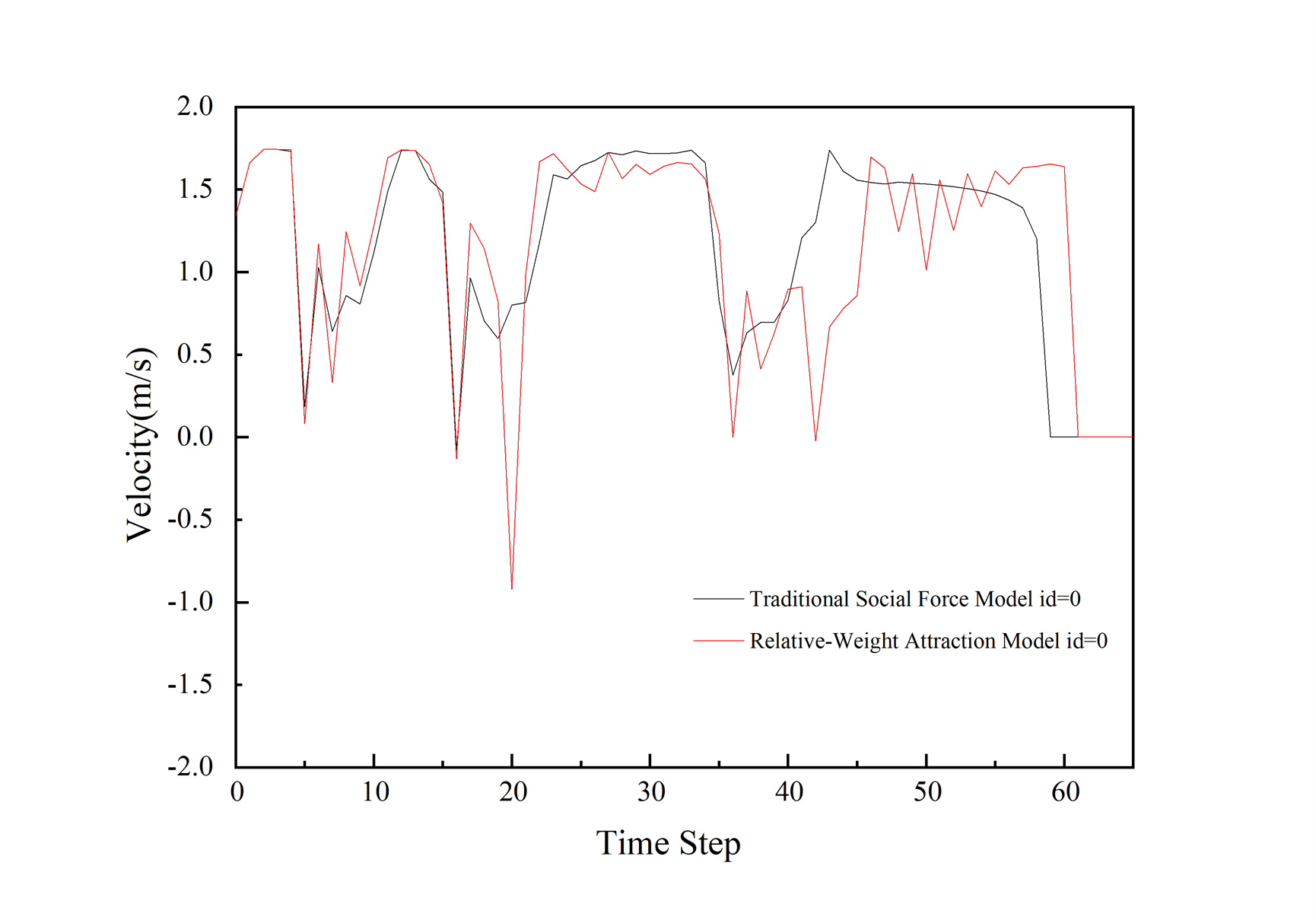}        
\caption{Velocity Simulation Line Chart Results for Pedestrian (ID=0) after Environmental Interaction  .}
\label{fig2} 
\end{figure*}

\begin{figure*}[h]
\centering
\includegraphics[scale=1]{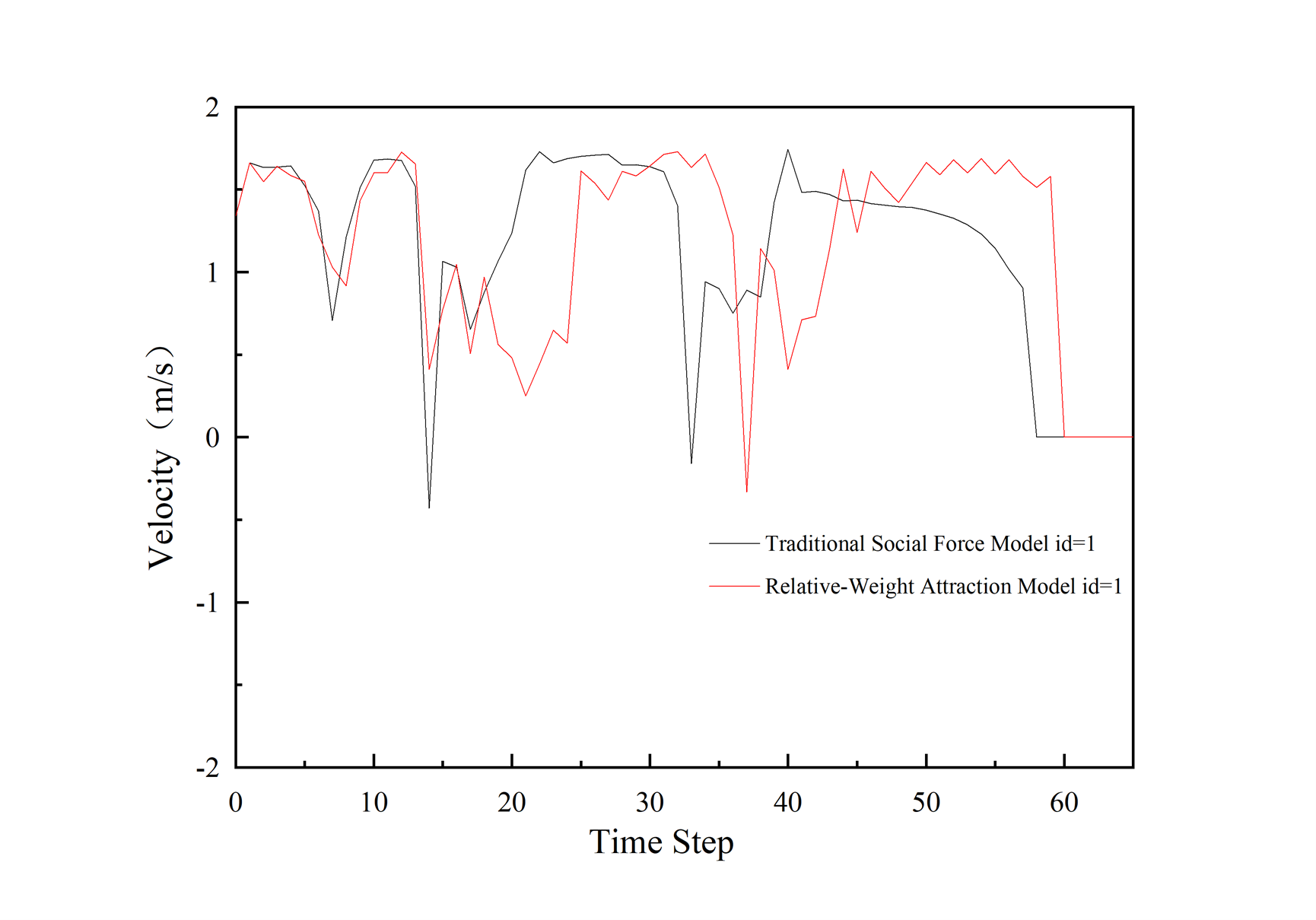}        
\caption{Velocity Simulation Line Chart Results for Pedestrian (ID=1) after Environmental Interaction .}
\label{fig2} 
\end{figure*}

\begin{figure*}[h]
\centering
\includegraphics[scale=1]{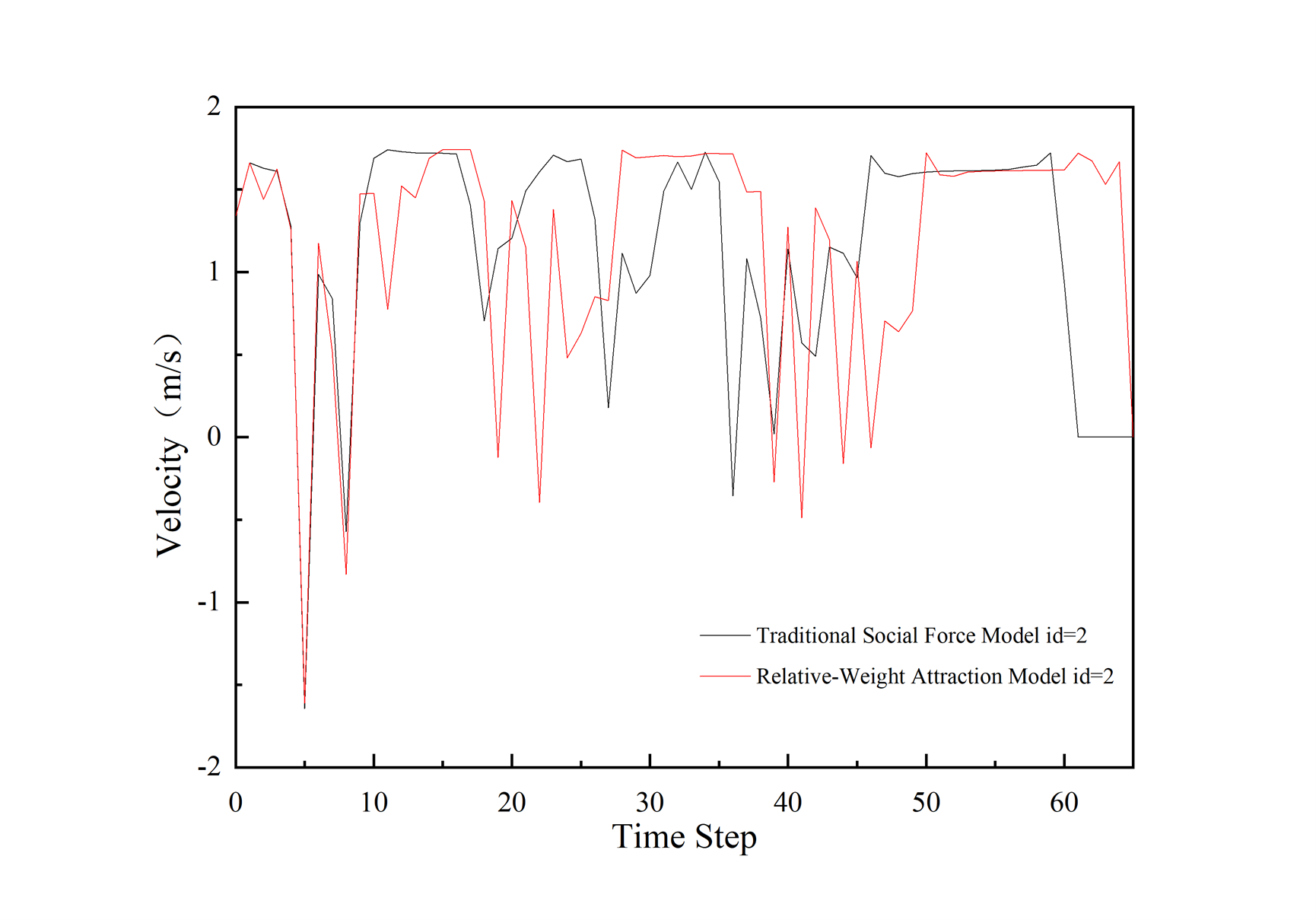}        
\caption{Velocity Simulation Line Chart Results for Pedestrian (ID=2) after Environmental Interaction .}
\label{fig2} 
\end{figure*}

\begin{figure*}[h]
\centering
\includegraphics[scale=1]{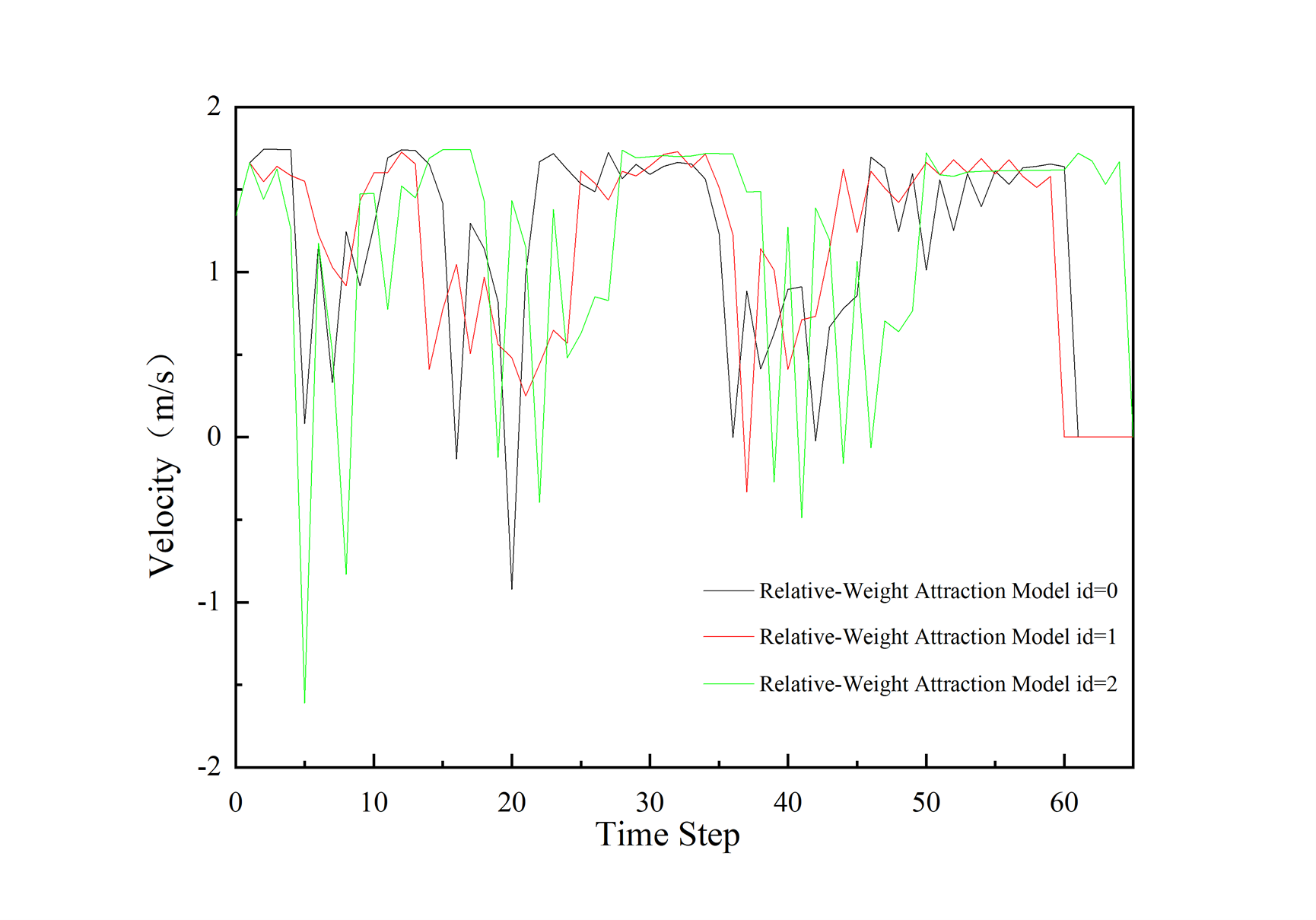}        
\caption{Comparative Simulated Line Chart Results of Velocity for Three Companion Pedestrians after Environmental Interaction under the Weighted Model .}
\label{fig2} 
\end{figure*}

In the context of obstacle collisions, the relative-weight attraction model incorporates social weights between leaders and members to emphasize group cohesion and coordination. Within obstacle-laden environments, leaders are required to guide the group to navigate around obstacles, while members must continuously adjust their positions and velocities to maintain the group structure. As evidenced in the figures, the weighted model exhibits more pronounced velocity fluctuations (e.g., red polylines for ID=0 and ID=2), indicating that members frequently adjust their speeds to synchronize with the leader’s movement rather than moving directly toward the target. In contrast, the traditional model relies primarily on simple repulsive and attractive forces, which may result in more direct pedestrian movement but often leads to group dispersion. This pattern aligns with the aforementioned small-group coordination behaviors observed in companion dynamics. In obstacle-laden environments, the evacuation speeds of all members in the relative-weight attraction model exhibit significant fluctuations, indicating that obstacles exert a substantial impact on group movement. The velocities frequently switch between positive and negative values, suggesting that the group is continuously navigating around obstacles or adjusting its path to avoid collisions

Analysis of the behavioral patterns of evacuation groups under the relative-weight model reveals that the leader (ID=0) exhibits three distinct deceleration events (approximately at time steps 10, 20, and 40). These deceleration events indicate that the leader encounters obstacles blocking its path. Simultaneously, the velocity profiles demonstrate that the followers (ID=1 and ID=2) exhibit highly synchronized velocity changes with the leader, displaying similar deceleration patterns at the same time steps. This synchronization indicates that the group maintains tight coordination, validating the effectiveness of the relative-weight model in simulating real-world companion behavior.

\subsection{4.4. Evaluation of the Improvement Effects in the Relative-Weight Attraction Model}
Based on multiple simulation data, a comparative analysis of the relative-weight attraction model and the traditional model was conducted from the dimensions of evacuation efficiency and pedestrian movement dynamics. Through quantitative analysis and statistical testing, the improvements and limitations of the enhanced model in terms of simulation accuracy, computational efficiency, and practical application value were explicitly evaluated. The animated simulation effects illustrating the evacuation of companion groups with and without leaders in a bidirectional passage, as well as the comparative impact of different companion ratios on evacuation efficiency relative to total population size, are demonstrated in Figure 10, Figure 11, and Figure 12.

\begin{figure*}[h]
\centering
\includegraphics[scale=1]{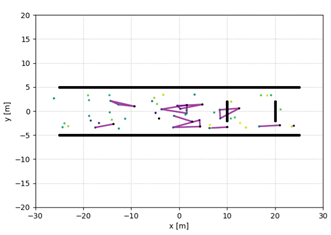}        
\caption{Evacuation Animation Schematic of Companion Groups with Leader Presence in a Bidirectional Passage.}
\label{fig2} 
\end{figure*}

\begin{figure*}[h]
\centering
\includegraphics[scale=1]{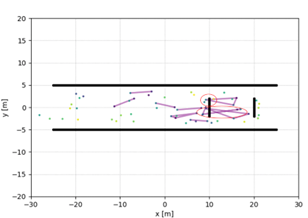}        
\caption{Evacuation Animation Schematic of Companion Groups without Leader Presence in a Bidirectional Passage.}
\label{fig2} 
\end{figure*}

Figures 10 and 11 present output animations from one of the 25 simulation runs. The point distribution reveals that companion groups exhibit a "V-shaped" or chain-like structure, which aligns with common patterns observed in real-world companion crowds. The leader (black dot) is positioned at the forefront of the group, guiding path selection and directing members toward the target. In the traditional social force model, evacuation of companion groups often encounters issues such as internal dispersion and followers’ difficulty in overcoming obstacles. As shown in Figure 11, the traditional model primarily relies on simple physical forces (e.g., repulsive and attractive forces) and lacks dynamic social weighting mechanisms. This results in insufficient attraction among individuals, causing group members to lose cohesion and act independently in obstacle-rich or crowded environments. The elongated purple lines in the figure reflect this dispersion tendency. Simultaneously, followers—without clear leadership—often hesitate, collide, or select suboptimal paths when encountering obstacles, leading to evacuation delays. Figure 11 shows crowd aggregation near obstacles, and subsequent animations reveal persistent difficulties in obstacle negotiation. In contrast, the weighted model effectively addresses these limitations by incorporating social weights and leadership roles, thereby enhancing group coordination and evacuation efficiency.

\begin{figure*}[h]
\centering
\includegraphics[scale=1]{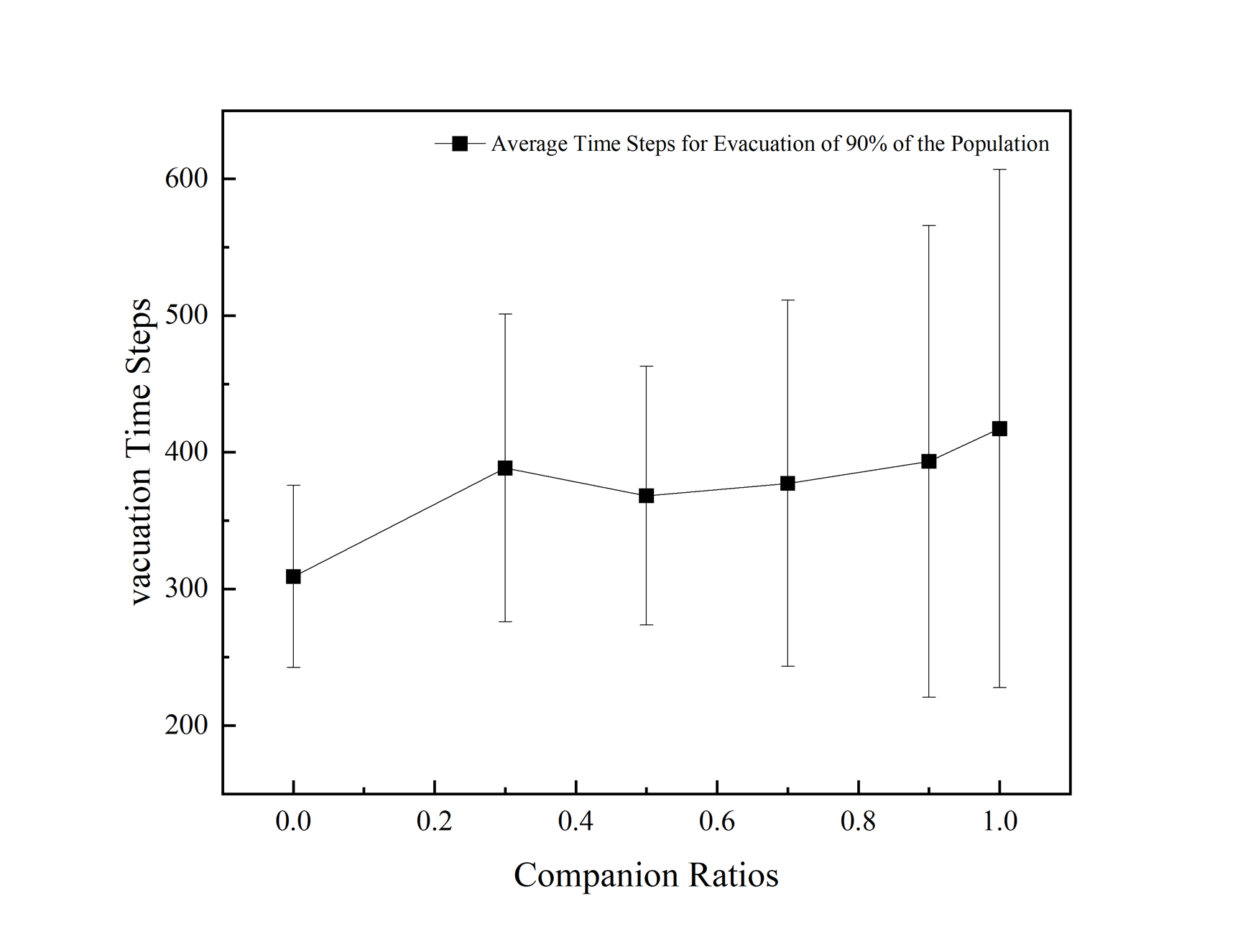}        
\caption{Comparative Results of Evacuation Efficiency versus Total Evacuation Population under Different Companion Ratios.}
\label{fig2} 
\end{figure*}

In the line chart, all curves exhibit an overall increasing trend in evacuation time steps as the total number of pedestrians rises, primarily attributable to channel congestion and decision-making delays caused by higher pedestrian density. When the aggregate pedestrian count reaches elevated levels (180–240 individuals), the growth rate of evacuation time steps decelerates, a phenomenon resulting from the saturation of channel capacity.
At a 50\% companion ratio, the relative-weight model generally outperforms the traditional model, indicating that the leader mechanism effectively enhances evacuation efficiency and reduces temporal delays in coordination under low pedestrian density conditions. However, in high-population scenarios (e.g., total population of 210 with 50\% companion ratio), its performance deteriorates across most companion ratios. This suggests that in high-density environments, leaders may struggle to guide group members effectively, and inter-group path conflicts could further exacerbate channel congestion.

\subsection{4.5. Impact of Varying Companion Ratios on Evacuation Efficiency}
The evacuation efficiency of personnel under varying companion ratios was investigated using the enhanced model, with the summarized results from 25 simulation runs and the corresponding efficiency trend line illustrated in Table 5 and Figure 13, respectively.

\begin{table*}[h]
\small\sf\centering
\caption{Experimental Scheme 3.\label{T1}}
\resizebox{\textwidth}{!}{
\begin{tabular}{lllllll}
\toprule
Companion Ratio&Average 90\% Evacuation Time Steps&Minimum 90\% Evacuation Time Steps&Maximum 90\% Evacuation Time Steps&Standard Deviation of 90\% Time Steps&Number of Simulation Runs&Valid Simulation Runs\\
\midrule
\texttt{0}&309.09 &188&450&65.30&25&23\\
\texttt{30\%}&388.43 &238&695&110.20&25&23\\
\texttt{50\%}&368.13 &241&721&92.55&25&23\\
\texttt{70\%}&377.21 &261&788&130.47&25&19\\
\texttt{90\%}&393.33 &220&982&168.52&25&21\\
\texttt{100\%}&417.29 &238&914&185.62&25&24\\
\bottomrule
\end{tabular}}
\end{table*}

\begin{figure*}[h]
\centering
\includegraphics[scale=1]{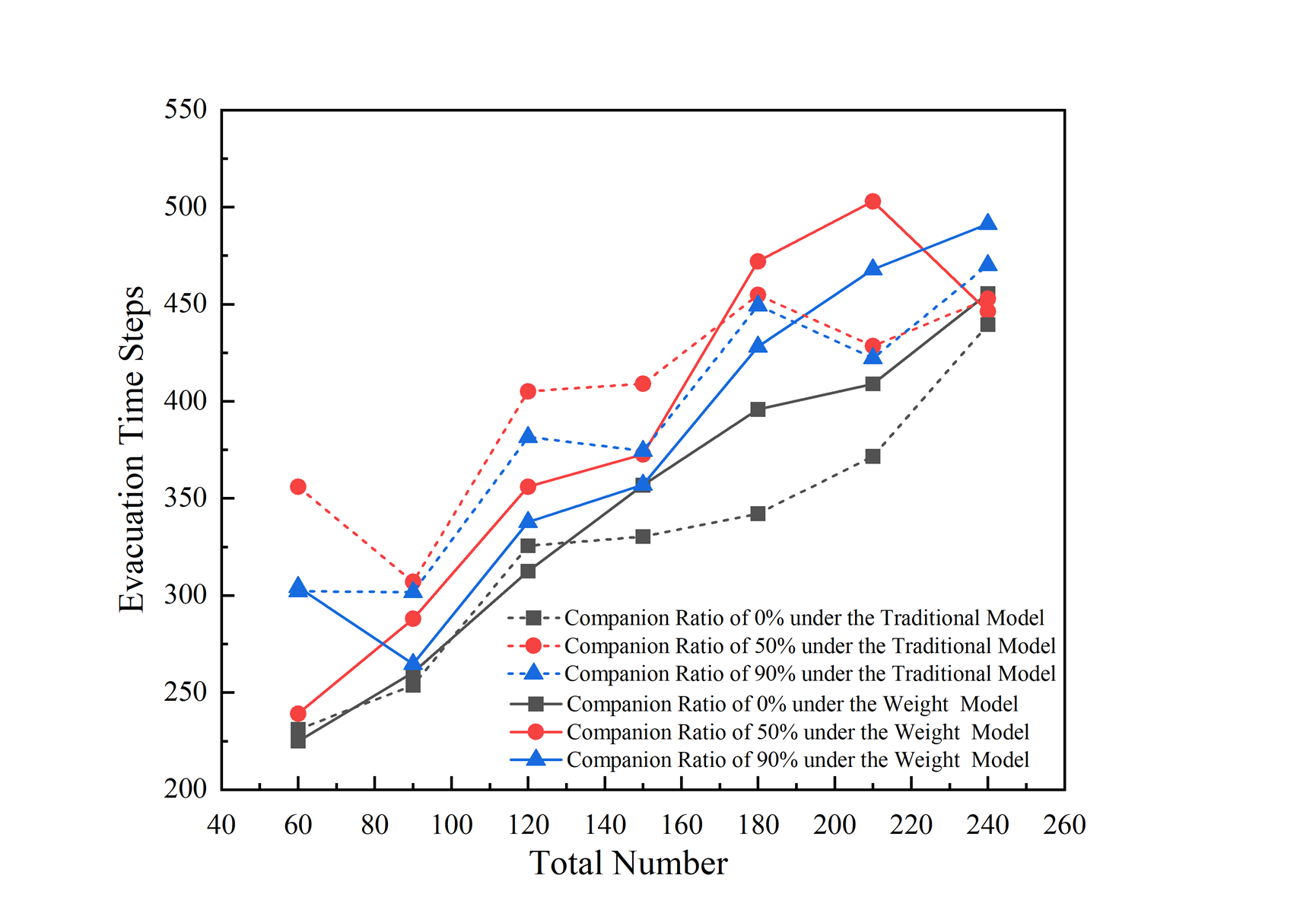}        
\caption{Evacuation Efficiency Results under Different Companion Ratios.}
\label{fig2} 
\end{figure*}

As evidenced by the line chart, the impact of companion ratios on evacuation efficiency exhibits a non-linear relationship, with evacuation time steps initially increasing, then decreasing, and subsequently rising again as the companion ratio escalates. The peak evacuation time occurs at a 30\% companion ratio, followed by a slight improvement at 50\%, before continuing to ascend. The error bars (representing standard deviation) lengthen with increasing companion ratios, signifying heightened variability in results and reduced reliability in predictive accuracy.

At a 50\% companion ratio, evacuation efficiency remains relatively high (averaging 368.13 time steps), indicating that moderate companionship combined with a leadership mechanism balances coordination costs with group benefits. However, even at this ratio, efficiency remains lower than that of 0\% companionship (independent individuals), suggesting that companion behavior generally reduces evacuation efficiency. Simultaneously, the weighted model performs poorly at high ratios (70\%–100\%), demonstrating that while leaders can reduce intra-group decision-making delays, they cannot fully overcome congestion and inter-group conflicts in high-density scenarios. This observation aligns with the conclusions from Scheme 2 above. Consequently, in emergency evacuation design, high companion ratios ($>$70\%) should be avoided. If companionship is necessary, maintaining a ratio of approximately 50\% is advisable, along with clearly defining leadership roles to optimize intra-group coordination. Passage design should minimize obstacle interference to prevent inter-group conflicts.

\section{5. Conclusion}
This study introduces a key enhancement to the classical social force model by incorporating an asymmetric relative-weight attraction mechanism, aiming to more accurately simulate the complex dynamics of companion groups with leaders during evacuation processes. The core innovation of the model lies in the integration of social weighting and leadership roles, coupled with improvements to the self-driving forces of pedestrians and their decision-making algorithms for desired directions. Consequently, a more realistic "Relative-Weight Attraction Social Force Model Considering Companion Behavior" is constructed.

The main findings and contributions of this study are summarized as follows:

1. Model Effectiveness and Superiority: Through simulation implementations programmed in Python and comparative analysis with the traditional social force model utilizing molecular potential (forces), the rationality and superiority of the proposed model in simulating companion groups were validated. This model effectively captures the enhanced regulation and followability of companion groups when encountering pedestrian collisions or obstacles, ensuring group cohesion and integrity in complex evacuation environments without easy dispersion.

2. Nonlinear Impact of Companion Ratios on Evacuation Efficiency: The study reveals a nonlinear relationship where evacuation time steps initially increase, then decrease, and subsequently rise again as the companion ratio escalates. Evacuation efficiency remains relatively high at a 50\% companion ratio. Moderate companionship combined with a leadership mechanism balances coordination costs and group benefits; however, collective companion behavior generally reduces overall evacuation efficiency compared to individual movement.

3. Model Applicability: The model is particularly suitable for simulating evacuation behaviors of companion groups with leaders (e.g., friends, colleagues, families). In such groups, leaders emerge naturally and guide members, and the model accurately captures this socio-dynamic process, providing valuable insights for emergency management and spatial design in these scenarios.


\begin{thebibliography}{21}
\providecommand{\natexlab}[1]{#1}
\providecommand{\url}[1]{\texttt{#1}}
\providecommand{\urlprefix}{URL }
\expandafter\ifx\csname urlstyle\endcsname\relax
  \providecommand{\doi}[1]{DOI:\discretionary{}{}{}#1}\else
  \providecommand{\doi}{DOI:\discretionary{}{}{}\begingroup \urlstyle{rm}\Url}\fi

\bibitem[{Couzin et~al.(2002)Couzin, Krause, James, Ruxton and Franks}]{R17}
Couzin ID, Krause J, James R, Ruxton GD and Franks NR (2002) Collective memory and spatial sorting in animal groups.
\newblock \emph{Journal of theoretical biology} 218(1): 1--11.

\bibitem[{Fridman et~al.(2012)Fridman, Zilka and Kaminka}]{R10}
Fridman N, Zilka A and Kaminka GA (2012) The impact of cultural differences on crowd dynamics.
\newblock In: \emph{AAMAS}. pp. 1343--1344.

\bibitem[{Fu et~al.(2014)Fu, Song, Lv and Lo}]{R18}
Fu L, Song W, Lv W and Lo S (2014) Simulation of emotional contagion using modified sir model: A cellular automaton approach.
\newblock \emph{Physica A: Statistical Mechanics and its Applications} 405: 380--391.

\bibitem[{Hall and Matsumoto(2004)}]{R19}
Hall JA and Matsumoto D (2004) Gender differences in judgments of multiple emotions from facial expressions.
\newblock \emph{Emotion} 4(2): 201.

\bibitem[{Han and Liu(2017)}]{R16}
Han Y and Liu H (2017) Modified social force model based on information transmission toward crowd evacuation simulation.
\newblock \emph{Physica A: Statistical Mechanics and its Applications} 469: 499--509.

\bibitem[{Helbing et~al.(2005)Helbing, Buzna, Johansson and Werner}]{R4}
Helbing D, Buzna L, Johansson A and Werner T (2005) Self-organized pedestrian crowd dynamics: Experiments, simulations, and design solutions.
\newblock \emph{Transportation science} 39(1): 1--24.

\bibitem[{Helbing et~al.(2000)Helbing, Farkas and Vicsek}]{R2}
Helbing D, Farkas I and Vicsek T (2000) Simulating dynamical features of escape panic.
\newblock \emph{Nature} 407(6803): 487--490.

\bibitem[{Helbing and Molnar(1995)}]{R1}
Helbing D and Molnar P (1995) Social force model for pedestrian dynamics.
\newblock \emph{Physical review E} 51(5): 4282.

\bibitem[{Hou et~al.(2014)Hou, Liu, Pan and Wang}]{R8}
Hou L, Liu JG, Pan X and Wang BH (2014) A social force evacuation model with the leadership effect.
\newblock \emph{Physica A: Statistical Mechanics and its Applications} 400: 93--99.

\bibitem[{Ivo et~al.(2021)Ivo, Cavalcante-Neto and Vidal}]{R6}
Ivo D, Cavalcante-Neto J and Vidal C (2021) A model for flexible representation of social groups in crowd simulation.
\newblock \emph{Computers \& Graphics} 101: 7--22.

\bibitem[{Langston et~al.(2006)Langston, Masling and Asmar}]{R12}
Langston PA, Masling R and Asmar BN (2006) Crowd dynamics discrete element multi-circle model.
\newblock \emph{Safety Science} 44(5): 395--417.

\bibitem[{Motsch and Tadmor(2011)}]{R11}
Motsch S and Tadmor E (2011) A new model for self-organized dynamics and its flocking behavior.
\newblock \emph{Journal of Statistical Physics} 144(5): 923.

\bibitem[{Organization(2016)}]{R21}
Organization IM (2016) Guidelines for evacuation analysis for new and existing passenger ships.

\bibitem[{Qu et~al.(2014)Qu, Gao, Xiao and Li}]{R14}
Qu Y, Gao Z, Xiao Y and Li X (2014) Modeling the pedestrian’s movement and simulating evacuation dynamics on stairs.
\newblock \emph{Safety science} 70: 189--201.

\bibitem[{Treiber et~al.(2000)Treiber, Hennecke and Helbing}]{R3}
Treiber M, Hennecke A and Helbing D (2000) Congested traffic states in empirical observations and microscopic simulations.
\newblock \emph{Physical review E} 62(2): 1805.

\bibitem[{Wu et~al.(2021)Wu, Chen, Li, Liu and Zheng}]{R15}
Wu W, Chen M, Li J, Liu B and Zheng X (2021) An extended social force model via pedestrian heterogeneity affecting the self-driven force.
\newblock \emph{IEEE transactions on intelligent transportation systems} 23(7): 7974--7986.

\bibitem[{Wu et~al.(2020)Wu, Wang and Hu}]{R20}
Wu X, Wang W and Hu Z (2020) Study on emotion contagion model and intervention strategies of panic in subway stations under emergency.
\newblock \emph{Computer Engineering and Applications} 56(12): 265--272.

\bibitem[{Xie et~al.(2021)Xie, Lee, Li, Shi, Cao and Zhang}]{R5}
Xie W, Lee EWM, Li T, Shi M, Cao R and Zhang Y (2021) A study of group effects in pedestrian crowd evacuation: Experiments, modelling and simulation.
\newblock \emph{Safety Science} 133: 105029.

\bibitem[{Yang et~al.(2023)}]{R9}
Yang S et~al. (2023) An improved social force model of emergency crowd evacuation considering the spread of mass panic in metro stations.
\newblock In: \emph{International Conference on Algorithms, High Performance Computing, and Artificial Intelligence (AHPCAI 2023)}, volume 12941. SPIE, pp. 1086--1092.

\bibitem[{Yuen and Lee(2012)}]{R13}
Yuen J and Lee E (2012) The effect of overtaking behavior on unidirectional pedestrian flow.
\newblock \emph{Safety Science} 50(8): 1704--1714.

\bibitem[{Zhang et~al.(2023)Zhang, Qu and Han}]{R7}
Zhang Q, Qu J and Han Y (2023) Pedestrian small group behaviour and evacuation dynamics on metro station platform.
\newblock \emph{Journal of Rail Transport Planning \& Management} 26: 100387.

\end{thebibliography}
\end{document}